\pdfoutput=1
\documentclass[twocolumn,preprintnumbers]{revtex4}
\usepackage{amsmath}
\usepackage{subfigure}
\usepackage{graphicx,psfrag}
\usepackage{bm}
\usepackage{multirow}

\begin{document}
\preprint{LA-UR-09-00111}

\title{A Boundary Approximation Algorithm for Distributed Sensor Networks}

\author{Michael I. Ham}
\author{Marko A. Rodriguez}
\affiliation{Theoretical Division - Center for Nonlinear Studies, Los Alamos National Laboratory, Los Alamos, New Mexico 87545}

\begin{abstract}
We present an algorithm for boundary approximation in locally-linked sensor networks that communicate with a remote monitoring station. Delaunay triangulations and Voronoi diagrams are used to generate a sensor communication network and define boundary segments between sensors, respectively. The proposed algorithm reduces remote station communication by approximating boundaries via a decentralized computation executed within the sensor network. Moreover, the algorithm identifies boundaries based on differences between neighboring sensor readings, and not absolute sensor values. An analysis of the bandwidth consumption of the algorithm is presented and compared to two naive approaches. The proposed algorithm reduces the amount of remote communication (compared to the naive approaches) and becomes increasingly useful in networks with more nodes.
\end{abstract}

\maketitle{}

\section{Introduction}

Sensor networks consist of a set of sensor devices that communicate with each other through a wired or wireless communication network. Such systems often communicate with a remote station and are a promising technology for remotely detecting physical phenomena such as forest fires, chemical leaks, or radioactive clouds. For many  applications, it is necessary that the network not only identify a phenomenon, but also determine the boundary of the detected phenomenon. For example, by establishing the boundary of a forest fire, a sensor network can help fire fighters determine where to concentrate their efforts. 

Adding more sensors to a network increases the accuracy of any boundary approximation algorithm, but consequently, increases the amount of data generated. Therefore, if all data is processed at the remote station, the required bandwidth is proportional to the size of the network. On the other hand, if only the nodes that sense the phenomenon report back, the required bandwidth is proportional to the size of the phenomenon. As an alternative to both of these naive approaches, we present a decentralized algorithm for boundary identification that limits remote station communication by determining the boundary segments of a phenomenon via a distributed computation that is carried out within the sensor network. Moreover, only sensors that identify a boundary ultimately communicate with the remote station. Therefore, the amount of remote communication is proportional to the size of the phenomenon's boundary.

\section{Two Naive Boundary Approximation Algorithms}

A naive solution to boundary approximation would be for each sensor to report its internal state to the remote station. Once the state of each sensor reaches the remote station, the station calculates the boundary using either a centralized version of the proposed method (to follow) or any other known centralized algorithm. Assuming $n$ nodes in the sensor network and each sensor has a cost of $\beta$ for a long-range transmission, the cost of this approach is $n\beta$. However, a complication with this approach is that the remote station must have the capacity to receive information from each node simultaneously in order to ensure an accurate snapshot of the phenomenon's location. Regardless, as the size of network increases, this approach becomes prohibitive since the cost scales with the number of sensors.

A second naive solution would be for only those sensors that detect the phenomenon to report back to the remote station. If $m$ is the number of nodes sensing a phenomenon, where $m \leq n$, then $m\beta$ would be cost of this algorithm. Using this method, remote communication scales with the size of the phenomenon, not the size of the network. A simple real-world example demonstrates a potential fault with this approach for certain phenomena. Imagine a sensor network whose function is to sense a gray-scale environment.  Given a constant light source, a sensation threshold can be determined. In such cases, only those sensors that sense a high enough gray-scale value would report to the remote station. However, once the light source is reduced beyond the preset sensation threshold, the phenomenon is no longer detected even though there exists a relative difference in the readings of the sensors at the boundary. Using the algorithm presented next, a relative analysis determines the boundary regardless of the strength of the light source (assuming there exists more light than absolute dark). Sensors use local communication to detect a boundary by comparing neighboring measurements and only sensors that identify a boundary communicate with the remote station. Finally, the number of sensors reporting scales with the size of the phenomenon's boundary. In many cases, the area of a phenomenon is likely to be significantly larger than the boundary. 

\section{The Proposed Boundary Approximation Algorithm}

\begin{figure*}[ht!]
	\includegraphics[width=0.8\columnwidth]{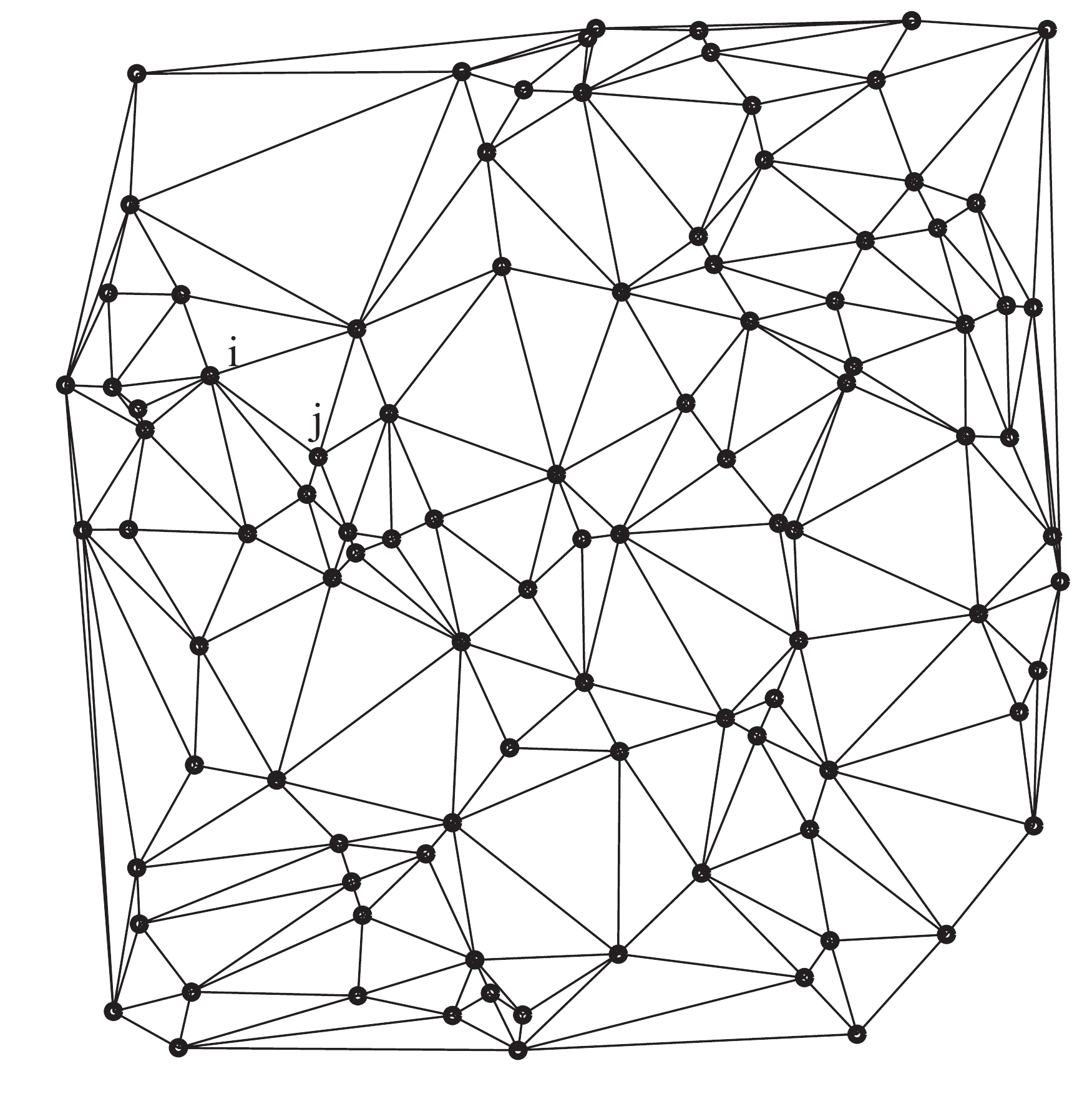}$\;\;\;\;\;\;$
	\includegraphics[width=0.8\columnwidth]{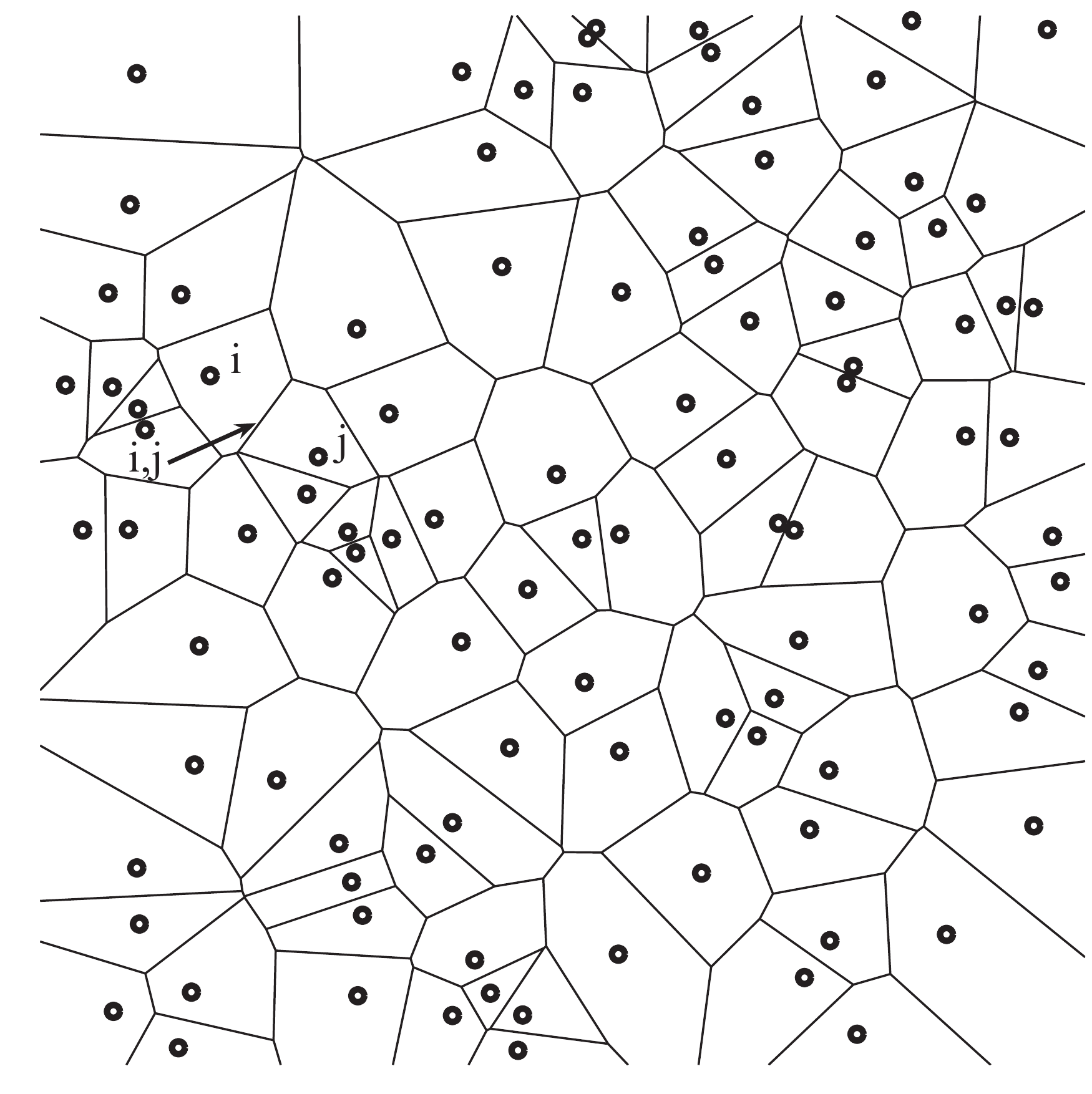}
	\caption{\label{fig:dv} \textbf{a.} A Delaunay triangulation and \textbf{b.} a Voronoi diagram for 100 randomly distributed sensors. Sensors $i$ and $j$ are identified as well as the $(i,j)$ Voronoi cell boundary that is equidistant between $i$ and $j$.}
\end{figure*}

Given a sensor network with $n$ nodes, a Delaunay triangulation is used to determine the neighbors of each node in the network \cite{delaunay:lee1980}. Next, a Voronoi diagram is generated to determine boundary segments between neighboring sensors. Such diagrams create cells with boundaries (segments), where all points on the cell boundary are equidistant between the two neighboring sensor nodes \cite{voronoi:aurenhammer1991}. Figure \ref{fig:dv} presents a visualization of a Delaunay triangulation (Figure \ref{fig:dv}a) and Voronoi diagram (Figure \ref{fig:dv}b) for $100$ randomly distributed nodes within a 2D space. Because there exists no distributed algorithm for calculating a Delaunay triangularization and a Voronoi diagram \cite{cover:li2003}, the sensor network's remote station can be used to calculate these. This one time calculation occurs only after sensors have been distributed and assumes that the remote station knows the exact location of each sensor.

Assume that a given sensor $i$ takes a measurement $\psi_i \in [0,1]$. In order to accomplish a local, distributed calculation of a phenomenon's boundary, $i$ must communicate with each of its neighbors  and compare its measurement with the measurements taken by those neighbors. If a particular neighbor $j$ of $i$ has a $\psi_j$ that is significantly different than $\psi_i$, then $i$ can assume that the phenomenon's boundary exists somewhere between itself and $j$. This threshold of difference is defined by $\theta \in [0,1]$ and a boundary exists when $\psi_i - \psi_j > \theta$. Since sensors have spatial gaps between them, the location of the boundary cannot be known exactly. The best approximation of the phenomenon's boundary is determined to be the line directly equidistant from $i$ and $j$. Conveniently, this line is the segment $(i,j)$ as defined by the Voronoi diagram. Therefore, once $(i,j)$ is determined to be a boundary segment, only this information needs to be transmitted to the remote station. Thus, only those sensors at the boundary of the phenomenon are communicating with the remote station. Moreover, the aggregate of all their reports is the approximated boundary.

\begin{figure*}[ht!]
	\includegraphics[width=0.8\columnwidth]{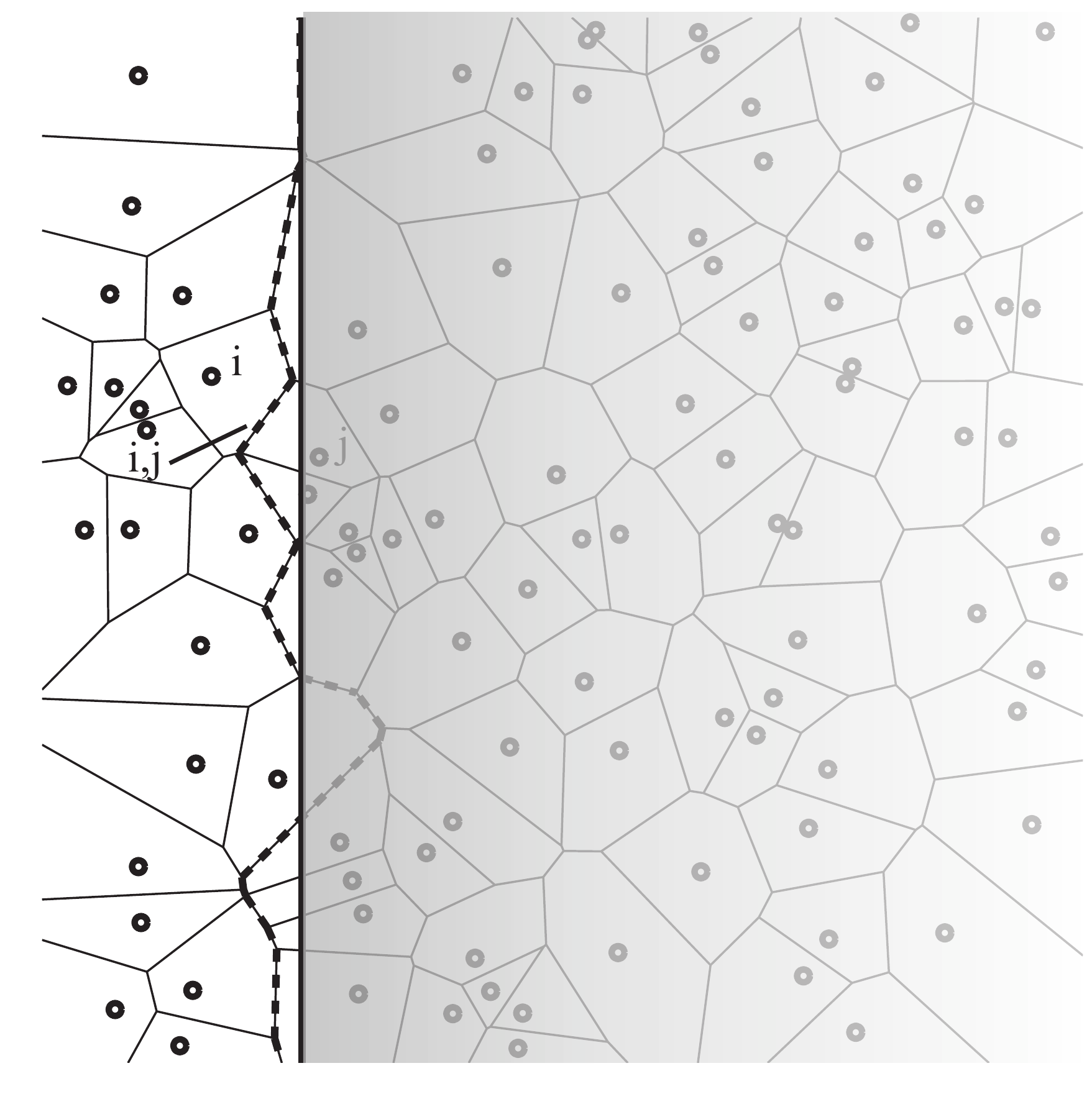}$\;\;\;\;\;\;$
	\includegraphics[width=0.8\columnwidth]{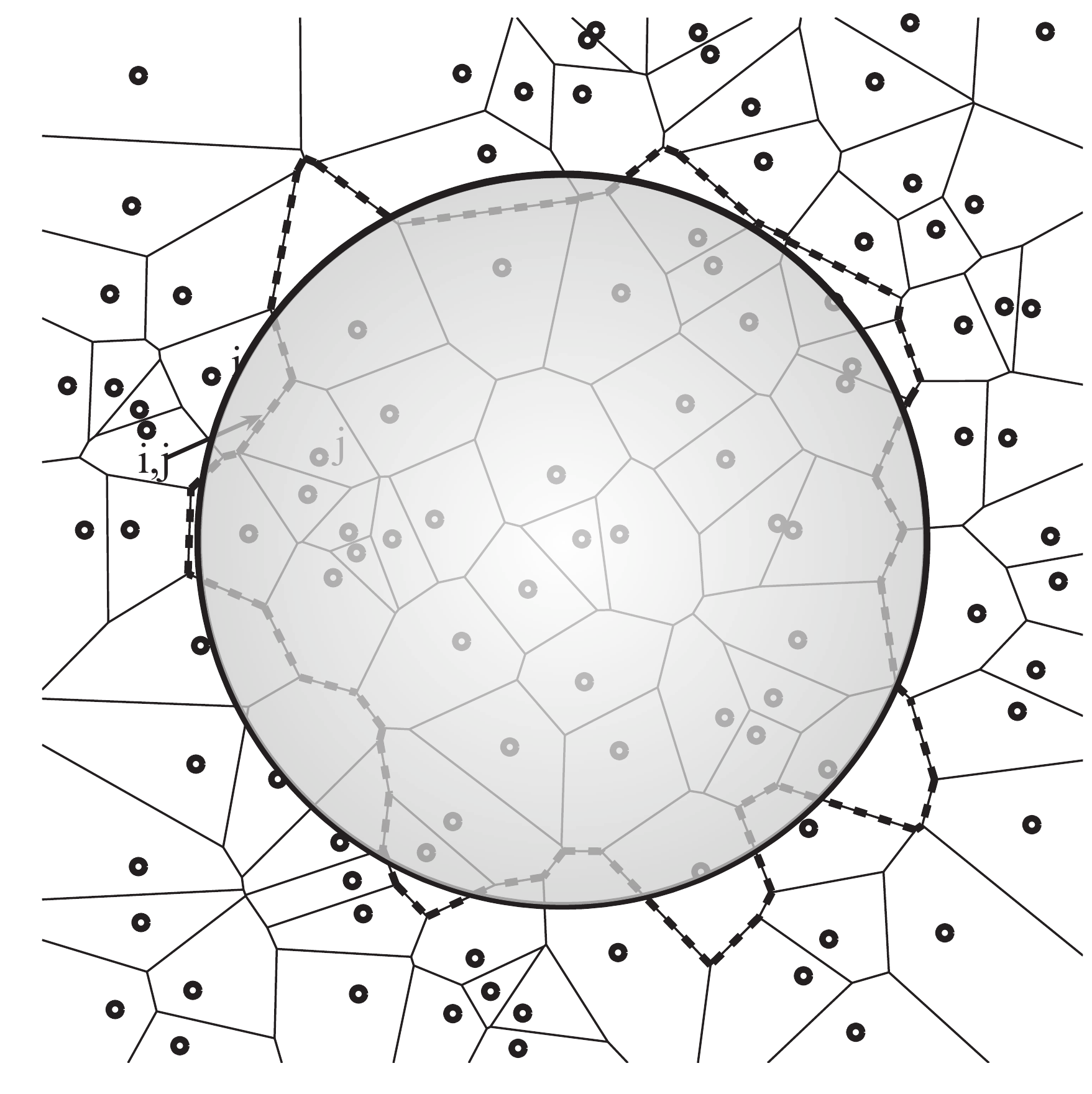}
	\caption{\label{fig:lc} A representation of the boundary approximated by the proposed algorithm for phenomena with \textbf{a.} linear and \textbf{b.} circular boundaries. Gray-scale shading denotes the phenomena. The boundary of the phenomena is the black line and the approximated boundary is the dashed line. The approximated boundary is always a collection of Voronoi cell segments.}
\end{figure*}

Figure \ref{fig:lc} presents two simulated phenomenon: one with a linear boundary and the other with a circular boundary. Each phenomena exist within a $100$ node sensor network. Table \ref{table:cost} presents the cost of each boundary detection approach for both phenomena, where $\epsilon$ denotes the relatively low cost of all inter-node communication \cite{comm:wierelthier2000}. It should be noted that for certain types of networks, especially radio wireless networks, the cost of local communication can be orders of magnitude less than long-range, remote communication. It is in these situations where the proposed algorithms is most efficient.

\begin{table}[h!]
\begin{footnotesize}
    \begin{tabular}{|c||c|c|c|} \hline
        boundary & 1$^\text{st}$ naive & 2$^\text{nd}$ naive & proposed \\
        \hline\hline
        linear & $100\beta$ & $80\beta$ & $11\beta + \epsilon$ \\
        circular & $100\beta$ & $38\beta$ & $24\beta + \epsilon$ \\\hline
    \end{tabular}
    \caption{\label{table:cost}The cost of each approach for simulated phenomena with linear and circular boundaries (see Figure \ref{fig:lc}). $\beta$ denotes the cost for remote communication and $\epsilon$ is the total cost for inter-network communication.}
\end{footnotesize}
\end{table}

\section{Monte Carlo Simulation}

\begin{figure*}[ht!]
    \includegraphics[width=0.75\textwidth]{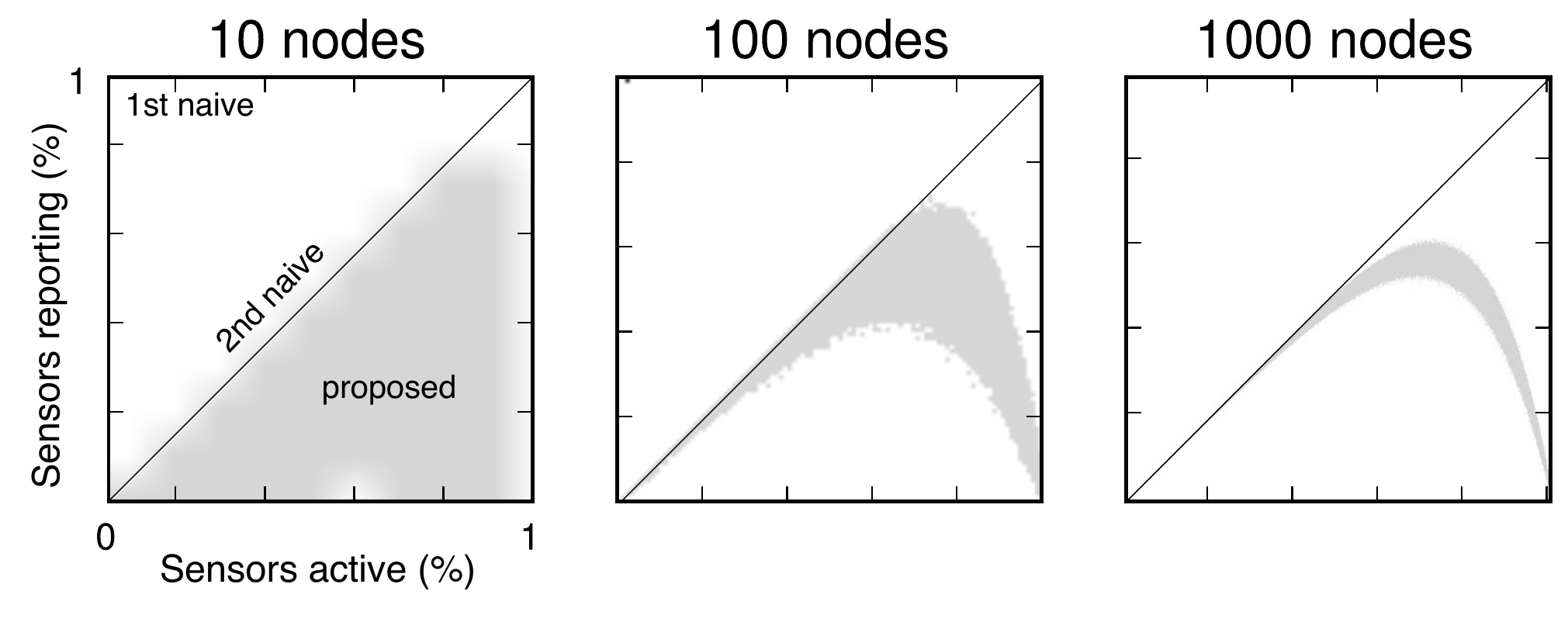}
    \caption{\label{fig:mont} A Monte Carlo simulation of the number of nodes (as a percent of the whole population) observing a phenomenon vs. the number (as a percent of the whole population) reporting back to the remote station in networks with $10$, $100$, and $1000$ nodes. For the proposed algorithm, as the number of nodes increases, the maximum number of nodes reporting to the remote station decreases.}
\end{figure*}

Monte Carlo simulations provide a means to test a system with many degrees of freedom, where an exhaustive parameter sweep is considered intractable \cite{metropolis_1949}. We utilize a Monte Carlo simulation of sensor networks containing $3$, $4$, $5$, $10$, $25$, $100$, $200$, $500$, and $1000$ nodes. For each population of nodes, one hundred different 2D space configurations are created within a fixed area. In each of the one hundred configurations, we activate a random set of nodes. The number of activated nodes is sequentially increased from $1$ to $n$. This random selection of nodes is done 100 times. For each resulting pattern, we calculate the number of sensors that would report back to the central station using the various boundary detection algorithms previously described. Figure \ref{fig:mont} presents the results for networks of $10$, $100$, and $1000$ nodes using the first naive approach (top horizontal line), the second naive approach (black diagonal line), and our proposed approach (gray cloud). Finally, Table \ref{table:t1} demonstrates for all networks tested, the maximum number of nodes reporting to the remote station. The results of Table \ref{table:t1} demonstrate that the proposed algorithm becomes more efficient as more nodes are added to the network. It should be noted, that for both naive approaches, the maximum number of reporting nodes is 100\%.

\begin{table}[h!]
\begin{footnotesize}
    \begin{tabular}{|c|c|}\hline
        number of nodes & max reporting (\%) \\\hline\hline
        3 & 100 \\
        4 & 100 \\
        10 & 90 \\
        25 & 84 \\
        100 & 72 \\
        200 & 68 \\
        500 & 63.6 \\
        1000 & 61.5 \\\hline
    \end{tabular}
    \caption{\label{table:t1}A Monte Carlo simulation identifies the maximum number of reporting nodes for the proposed algorithm.}
\end{footnotesize}
\end{table}

\section{Conclusions}

Related work on boundary approximation in sensor networks relies mainly on local communication and distributed computation \cite{edge:liao2003,local:chin2003,image:deva2003,bound:nowak2003}. However, most boundary approximation algorithms do not determine the boundary of the phenomenon, only the sensors that lie at the boundary. By knowing which sensor's lie at the boundary, the remote station can then estimate the actual line defining the boundary of the phenomenon. In contrast, the proposed algorithm computes the phenomenon's boundary internal to the network without reliance on the remote station. Local communication is used to identify pairs of nodes with readings whose difference is greater than $\theta$. One of the two nodes transmits the pair's Voronoi segment to the remote station. The aggregation of all these segments is the approximated boundary of the phenomenon.

It should be noted that a boundary can never be determined exactly since spatial gaps exist between sensors. Therefore, any calculation of the phenomenon's boundary is only an approximation. To reduce boundary location uncertainty, more sensors can be added to the network. As sensor networks increase in size, it is important to keep costs to a minimum. Costs can be reduced by utilizing low bandwidth communication and energy efficient processors with moderate clock speeds and small amounts of on-board memory. The proposed algorithm helps achieve one of these objectives by reducing remote station communication. The algorithm may prove useful in wireless sensor networks where radio communication over long distances requires significantly more energy than local communication \cite{comm:wierelthier2000}.

\section*{Acknowledgements}

We would like to thank Levi Larkey and Vadas Gintautas for their contributions to this article. This research was funded by a U.S. Department of Education GAANN Fellowship and an IC Postdoctoral Fellowship. Further support was provided by the Los Alamos National Laboratory.

\end{document}